\documentclass[draft,jgrga]{agutex}
\usepackage{amsmath} 
\usepackage{amssymb,color} 
\usepackage[]{graphicx}
\authorrunninghead{Doss et al.}
\titlerunninghead{Asymmetric Reconnection with Flow Shear}
\usepackage{verbatim}
\usepackage{mathrsfs}

\begin{document}

\title{Asymmetric magnetic reconnection with a flow shear and
  applications to the magnetopause}

\authors{C.~E.~Doss\altaffilmark{1}, C.~M.~Komar\altaffilmark{1,a},
  P.~A.~Cassak\altaffilmark{1}, F.~D.~Wilder\altaffilmark{2},
  S.~Eriksson\altaffilmark{2}, and J.~F.~Drake\altaffilmark{3}}
\altaffiltext{1}{Department of Physics and Astronomy, West Virginia
  University, Morgantown, West Virginia, USA}
\altaffiltext{2}{Laboratory of Atmospheric and Space Physics,
  University of Colorado, Boulder, Colorado, USA}
\altaffiltext{3}{Institute for Research in Electrons and Applied
  Physics, University of Maryland, College Park, Maryland, USA}
 \authoraddr{P.~A.~Cassak, Department of Physics
  and Astronomy, White Hall, Box 6315, West Virginia University,
  Morgantown, WV 26506. (Paul.Cassak@mail.wvu.edu)}

\begin{abstract}
  We perform a systematic theoretical and numerical study of
  anti-parallel two-dimensional magnetic reconnection with asymmetries
  in the plasma density and reconnecting magnetic field strength in
  addition to a bulk flow shear across the reconnection site in the
  plane of the reconnecting fields, which commonly occurs at planetary
  magnetospheres.  We analytically predict the speed at which an
  isolated X-line is convected by the flow, the reconnection rate, and
  the critical flow speed at which reconnection no longer takes place
  for arbitrary reconnecting magnetic field strengths, densities, and
  upstream flow speeds, and we confirm the results with two-fluid
  numerical simulations.  The predictions and simulation results
  counter the prevailing model of reconnection at Earth's dayside
  magnetopause which says reconnection occurs with a stationary X-line
  for sub-Alfv\'enic magnetosheath flow, reconnection occurs but the
  X-line convects for magnetosheath flows between the Alfv\'en speed
  and double the Alfv\'en speed, and reconnection does not occur for
  magnetosheath flows greater than double the Alfv\'en speed.  In
  particular, we find that X-line motion is governed by momentum
  conservation from the upstream flows, which are weighted differently
  in asymmetric systems, so the X-line convects for generic conditions
  including sub-Alfv\'enic upstream speeds.
  For the reconnection rate, as with symmetric reconnection, it drops
  with increasing flow shear and there is a cutoff speed above which
  reconnection is not predominant.  However, while the cutoff
  condition for symmetric reconnection is that the difference in flows
  on the two sides of the reconnection site is twice the Alfv\'en
  speed, we find asymmetries cause the cutoff speed for asymmetric
  reconnection to be higher than twice the asymmetric form of the
  Alfv\'en speed.  The stronger the asymmetries, the more the cutoff
  exceeds double the asymmetric Alfv\'en speed.  This is due to the
  fact that in asymmetric reconnection, the plasma with the smaller
  mass flux into the dissipation region contributes a smaller mass to
  the dissipation region, so the effect of its flow on opposing the
  release of energy by the reconnected magnetic fields is diminished
  and the reconnection is not suppressed to the extent previously
  thought.  The results compare favorably with an observation of
  reconnection at Earth's polar cusps during a period of northward
  interplanetary magnetic field, where reconnection occurs despite the
  magnetosheath flow speed being more than twice the magnetosheath
  Alfv\'en speed, the previously proposed suppression condition.
  These results are expected to be of broad importance for
  magnetospheric physics of Earth and other planets; particular
  applications are discussed.
\end{abstract}

\begin{article}
\section{Introduction}
\label{section::Introduction}

A key element controlling the interaction of the solar wind with
Earth's magnetosphere is the nature and efficiency of the magnetic
reconnection process at the dayside magnetopause.  During
reconnection, solar wind magnetic field lines effectively break and
cross-connect with terrestrial magnetic field lines.  As a
consequence, solar wind plasma is able to enter the magnetosphere and
the reconnected magnetic field lines convect tailward
\citep{Dungey61}.  This is a crucial aspect of space weather
phenomena.  The focus of this study is how the bulk flow of the solar
wind around the magnetosphere itself affects the dayside reconnection
process; we focus on (1) the convection of the reconnection site by
the bulk flow, (2) the effect on the reconnection rate, and (3) the
critical bulk flow speed above which reconnection is not the dominant
effect.

Bulk flow is expected to be most important when reconnection occurs
near the polar cusps.  This is because the bulk magnetosheath flow
around the magnetopause acquires a potentially sizable component
parallel or anti-parallel to the magnetic field when the reconnection
site is near the cusps.  This affects the reconnection site very
differently than when reconnection is near the subsolar point, where
the flow is predominantly not aligned with the reconnecting
magnetosheath magnetic field.  As pointed out by \citet{Dungey63},
magnetic reconnection is likely to occur near the polar cusps when the
interplanetary magnetic field (IMF) has a northward component.
Detections of high latitude reconnection are quite common
\citep{Gosling86,Gosling91,Gosling96,Kessel96,Fuselier00b,Onsager01,Avanov01,Federov01,Phan01,Frey03,Phan03,Lavraud04,Lavraud05,Retino05,Retino06,Phan06,Phan07,Hasegawa08,Fuselier10,Fuselier12,Fuselier14b,Muzamil14,Wilder14}.

One reason the effect of the solar wind flow on reconnection is
interesting is that there have been conflicting results on whether the
cusp reconnection site is stationary or is convected tailward.
\citet{Cowley89} and \citet{Gosling91} suggested that in order to see
sunward flow from a reconnection event poleward of the cusp with a
stationary X-line (where the reconnecting magnetic field goes to
zero), the magnetosheath flow speed should be sub-Alfv\'enic; if the
flow is super-Alfv\'enic, the X-line would have to convect tailward.
If the magnetosheath flow is more than double the magnetosheath
Alfv\'en speed, reconnection could not occur.  Tailward convection of
an X-line has been seen in global magnetospheric simulations using
magnetohydrodynamic simulations of Earth \citep{Berchem95} and hybrid
(kinetic ions with fluid electrons) simulations of Mercury
\citep{Omidi06}.  In the latter, once the X-line convected far enough
tailward, a new X-line formed.  This behavior was identified in
Cluster observations \citep{Hasegawa08}.  In other studies, the
stability of auroral signals associated with high-latitude
reconnection suggest reconnection sites remain stationary
\citep{Fuselier00b,Frey03}, although there are uncertainties about
whether a lack of change in auroral signatures necessarily precludes
repeated X-line generation and tailward convection.  It was also
suggested that reconnection with magnetosheath flow speed exceeding
twice the magnetosheath Alfv\'en speed can occur because suppressing
reconnection would introduce a pileup of magnetic flux that creates
plasma depletion layers, which increase the local Alfv\'en speed
\citep{Fuselier00}.

Another reason this topic is interesting is that a flow shear, such as
that caused by the solar wind, slows down the reconnection process and
can even stop it.  It has been shown analytically and numerically
using the magnetohydrodynamic (MHD) model for symmetric systems (with
equal and opposite reconnecting magnetic fields and the same mass
density on either side of the reconnection site) that a
super-Alfv\'enic flow shear completely suppresses reconnection, while
reconnection still occurs for sub-Alfv\'enic flow shear
\citep{Mitchell78,Chen90,LaBelleHamer94}.  When reconnection occurs
with a sub-Alfv\'enic flow shear present, there is a decrease in the
reconnection rate
\citep{Chen97,Li10,Faganello10,Cassak11a,Voslion11,Zhang11,Wu13} and
outflow speed \citep{Cassak11b}.  The situation is more complicated in
more realistic models than MHD; there are regimes in the Hall-MHD
model in which both tearing (the linear form of reconnection) and
Kelvin-Helmholtz can be simultaneously linearly unstable
\citep{Chacon03}.

The suppression of reconnection by a flow shear is potentially of
broader importance.  It was suggested that suppression of reconnection
by flow shear limits the length of the X-line ({\it i.e.,} the
separator) at the dayside \citep{Borovsky13a,Borovsky13b}.  However,
\citet{Komar15} pointed out reconnection suppression via flow shear is
not expected to play a role for southward IMF orientations because the
bulk flow is oriented out of the reconnection plane (along the flanks)
rather than parallel to the reconnecting magnetic field (toward the
poles), although it does locally decrease the reconnection rate when
the IMF is directed northward.

The effect of flow shear on reconnection is also expected to be
relevant at other planets.  Earth's solar wind-magnetospheric
interaction is qualitatively different from Jupiter's and Saturn's,
where the planet's relatively rapid rotation contributes to the global
convection pattern \citep{Vasyliunas83}.  Studies have investigated
the extent that reconnection occurs at the outer planets'
magnetopauses and whether flow shear plays a role in preventing it
\citep{Masters12,Desroche12,Desroche13,Masters14,Fuselier14}.  The
solar wind-magnetospheric interaction at Mercury is similar to Earth's
but on a much more rapid time scale \citep{Slavin09}, so flow shear
may also affect reconnection at Mercury.

For applications at Earth's magnetosphere, it is important to note
that the magnetospheric magnetic field is typically a few times
stronger than the magnetosheath magnetic field and the solar wind
plasma in the magnetosheath typically has a much higher density than
that of the magnetosphere \citep{Phan96,Ku97}, {\it i.e.,} the
reconnection is asymmetric.  Consequently, it is crucial to extend
studies of flow shear to asymmetric systems.  There has been much work
of late on asymmetric reconnection in the absence of flow shear; we
summarize only those results most germane to the present study.  The
rate of two-dimensional (2D) asymmetric reconnection (with
anti-parallel magnetic fields) has been studied
\citep{Borovsky07,Cassak07d}; it will be summarized in
Sec.~\ref{section::Theory}.  In addition, it was found that the X-line
and the stagnation point, where the inflowing plasma bulk flow speed
goes to zero, are generally not in the same location for asymmetric
reconnection \citep{Priest00b,Siscoe02,Dorelli04,Cassak07d}.

We know of only a few numerical studies of the impact of flow shear on
asymmetric magnetic reconnection.  \citet{LaBelleHamer95} studied
reconnection with an asymmetric density and a flow shear in fluid
simulations; they suggested that reconnection is suppressed if the
flow shear exceeds the Alfv\'en speed on either side of the layer.
\citet{Tanaka10} used kinetic particle-in-cell simulations to study
reconnection with a density asymmetry, a flow shear, and a guide
field; they observed that the X-line convects in the outflow direction
with contributions from both the flow shear and the diamagnetic drift
\citep{Swisdak03}.

In the present study, we use theoretical and numerical techniques to
study asymmetric reconnection with arbitrary upstream parallel flow
speeds.  We predict the bulk convection speed of an isolated X-line
using a simple fluid analysis.  An interesting conclusion is that the
asymmetries introduce qualitative differences compared to symmetric
reconnection.  In particular, if the upstream flow is equal and
opposite on the two sides, the X-line is stationary for symmetric
reconnection but convects for asymmetric reconnection.  We also
predict the reconnection rate of asymmetric reconnection with upstream
flow, including a condition for the critical upstream flow speed
required to suppress reconnection.  The critical flow shear (half the
difference of the flows on either side) for symmetric reconnection is
the Alfv\'en speed, so one might expect the cutoff for asymmetric
reconnection to be the asymmetric generalization of the Alfv\'en
speed.  However, we show that the cutoff exceeds the asymmetric
Alfv\'en speed, and the cutoff speed becomes much larger than the
asymmetric Alfv\'en speed when the asymmetry is large, such as the
typical conditions at Earth's magnetopause.  Consequently, an isolated
X-line at Earth's polar cusps would almost never be suppressed by flow
shear.  We use 2D two-fluid simulations to confirm the predictions.
(We point out that particle-in-cell simulations are ostensibly a
better tool to test the theory because of their more accurate
description of kinetic scale physics and plasma mixing in
collisionless plasmas, but fluid simulations are employed here as a
first step because they accurately portray the large scale physics
while being less noisy and therefore easier to compare to the new
theory presented here.  We employ this approach because it has proven
prudent for asymmetric reconnection without flow shear
\citep{Cassak07d,Cassak08b,Cassak09c,Malakit10}.  Important features
which are present in particle-in-cell simulations but not two-fluid
simulations are discussed in detail in the conclusion section.)  We
then show that the predictions are consistent with recent observations
of a cusp reconnection event for which the magnetosheath flow exceeded
twice the magnetosheath Alfv\'en speed.  The present results are in
stark contrast to the previous understanding of the effect of flow
shear on reconnection based on the \citet{Cowley89} results, as
X-lines convect even for sub-Alfv\'enic flow and reconnection occurs
for magnetosheath flow speeds much greater than the magnetosheath
Alfv\'en speed.

The layout of this paper is as follows: Section~\ref{section::Theory}
has the derivation of an expression for the convection speed of the
X-line and a prediction for the reconnection rate for asymmetric
reconnection with a flow shear.  Section~\ref{section::Setup} reviews
the numerical techniques and parameters for the simulations.
Section~\ref{section::Results} presents the results of our
simulations.  Section~\ref{section::applications} discusses
implications for observations and applications to planetary
magnetospheres.  Section~6 summarizes the results and discusses
limitations of the study.

\section{Theory}
\label{section::Theory}

We begin by defining system variables for asymmetric reconnection with
upstream flow parallel or anti-parallel to the reconnecting magnetic
fields.  The upstream magnetic fields above and below the dissipation
region are ${\bf B}_1$ and ${\bf B}_2$, which are assumed
anti-parallel ({\it i.e.,} there is no guide field).  The magnetic
fields are in opposite directions, so we will use $B_1$ and $B_2$ as
the magnitudes of the fields.  (Their direction does not impact the
present analysis as long as they are oppositely directed.)  The
upstream mass densities are $\rho_1$ and $\rho_2$, and the density in
the downstream region is $\rho_{{\rm out}}$.

The upstream flow speeds $v_{L,1}$ and $v_{L,2}$ are defined in a
stationary frame (with the planet in question at rest or, in the case
of a simulation, the rest frame of the simulation), where $L$ refers
to the reconnecting magnetic field direction (as in boundary normal
coordinates).  Each speed is defined as positive if to the right and
negative if to the left.  The convection speed of the X-line is
defined as $v_{{\rm drift}}$.  The inflow speeds (normal to the
dissipation region) are $v_{in,1}$ and $v_{in,2}$, and $L_d$ and
$\delta$ are the half-length and half-width of the dissipation region,
respectively.

\begin{figure}[t]
\centering
\noindent\includegraphics[width=20pc]{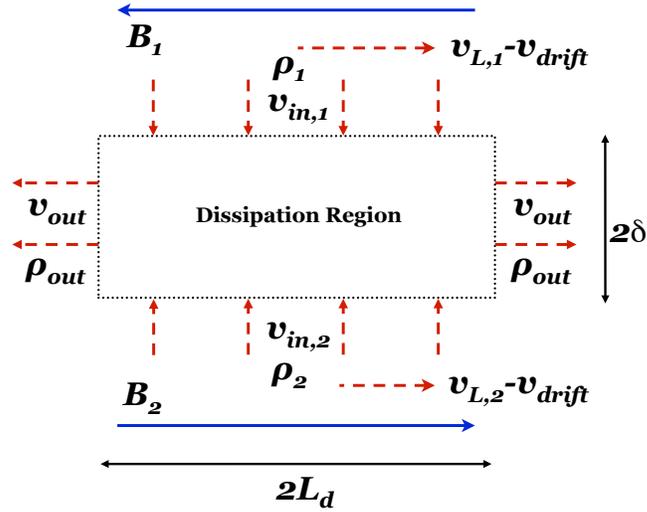}
\caption{Sketch of a reconnection region for asymmetric reconnection
  with flow shear in the reference frame the dissipation region.  Red
  dashed lines show components of plasma velocities and blue solid
  lines represent the reconnecting magnetic fields.}
\label{fig::Flux_Diagram}
\end{figure}

It is convenient to analyze this system in the rest frame of the
X-line, so we transform into the reference frame moving at a speed
$v_{{\rm drift}}$ relative to the stationary reference frame.  The
dissipation region in this reference frame is sketched in
Fig.~\ref{fig::Flux_Diagram}.  In this frame, the upstream parallel
flow speeds are given by $v_{L,1}-v_{{\rm drift}}$ and
$v_{L,2}-v_{{\rm drift}}$.  The outflow speed is $v_{{\rm out}}$,
which is expected to be the same in both outflow directions in the
rest frame of the X-line.  We define $x$ to be the direction of the
outflow ($L$ in boundary normal coordinates).  A number of potentially
important effects are ignored to keep the analysis tractable; these
are discussed in Sec.~6.

\subsection{Prediction of the X-line Convection Speed}
\label{section::speed}

We first consider the convection of an isolated X-line for 2D
anti-parallel asymmetric reconnection with arbitrary upstream parallel
flow speeds.  The physical reason that the X-line can convect in the
outflow direction is that the upstream plasmas carry momentum in the
outflow direction; by conservation of momentum, if the upstream plasma
has a non-zero net momentum, then the dissipation region will too.
This is true both for symmetric and asymmetric reconnection.  For
symmetric reconnection, the X-line is expected to convect at the
average of the upstream flows.  For asymmetric reconnection, the
inflow speed is different for the two upstream sides when the system
has asymmetric magnetic field strengths \citep{Cassak07d}, so the two
upstream plasmas do not contribute momentum in the outflow direction
equally.  Consequently, as we will show, the X-line convects in the
outflow direction for asymmetric reconnection even for equal and
opposite upstream flow speeds.  An alternate, but equivalent,
interpretation is that the X-line and stagnation point are separated
during asymmetric reconnection \citep{Cassak07d}, so the side with the
plasma that crosses the X-line imposes a flow in the outflow direction
at the X-line, causing it to convect.

To estimate the convection speed, we use a fluid description of the
plasma to find the bulk effects of the flow.  The governing equation
is the momentum equation of MHD, which in conservative form is
\begin{equation}
  \frac{\partial(\rho\textbf{v})}{\partial t} + \nabla \cdot 
  \left[ \rho\textbf{vv} + \left(P + \frac{B^2}{8\pi}\right)\textbf{I} - 
    \frac{\textbf{BB}}{4\pi} \right] = 0,
\label{eqn::cons_p}
\end{equation}
where $\rho$ is the plasma density, $\textbf{v}$ is the bulk velocity,
$P$ is the gas pressure, $\textbf{B}$ is the magnetic field, and
$\textbf{I}$ is the unit tensor.  Take the volume integral of
Eq.~(\ref{eqn::cons_p}) over the entire dissipation region.  For
steady-state reconnection, the time derivative term vanishes.  From
the divergence theorem, the remaining term becomes
\begin{equation}
  \oint d\textbf{S}\cdot\left[\rho\textbf{vv} + \left(P + 
      \frac{B^2}{8\pi}\right)\textbf{I} - \frac{\textbf{BB}}{4\pi} \right] = 0,
\label{eqn::steady_int}
\end{equation}
where $d\textbf{S}$ is a differential area element directed normal to
the boundary of the dissipation region.  For both the upstream and
downstream boundaries, the magnetic field is approximately parallel to
the boundaries, so the $\textbf{BB}$ term does not contribute.  Total
pressure is balanced across the dissipation region, so the pressure
term has no net contribution.  The $x$ component of the surviving term
of Eq.~(\ref{eqn::steady_int}) is
\begin{equation}
  \oint d\textbf{S}\cdot\left(\rho\textbf{v}v_x\right) = 0.
\label{eqn::surface_int}
\end{equation}
Performing a scaling analysis on this gives
\begin{equation}
  2L_d \rho_1 [v_{in,1} (v_{L,1} - v_{{\rm drift}})] + 2L_d \rho_2 [v_{in,2} 
  (v_{L,2} - v_{{\rm drift}})] \sim 0,
\label{eqn::surface_scale}
\end{equation}
where the two terms are the contributions from the two upstream sides
and the $x$ directed momentum flux on the downstream edges cancel.
Conservation of magnetic flux implies $v_{in,1} B_{1} \sim v_{in,2}
B_2$ \citep{Cassak07d}, so solving for $v_{{\rm drift}}$ gives
\begin{equation}
  v_{{\rm drift}}\sim\frac{\rho_1B_2v_{L,1}+\rho_2B_1v_{L,2}}
{\rho_1B_2+\rho_2B_1}.
\label{eqn::drift_eqn}
\end{equation}
This gives the convection speed of the X-line in the outflow direction
for arbitrary upstream densities, reconnecting magnetic field
strengths, and parallel flow speeds.  As discussed earlier, $v_{{\rm
    drift}}$ is non-zero for symmetric anti-parallel flow ($v_{L,1} =
- v_{L,2}$) when the magnetic fields and/or densities are asymmetric.
This shows that the X-line convects regardless of upstream flow
speeds, not only for super Alfv\'enic flow as suggested in
\citet{Cowley89}.

We consider particular limits of this expression.  For any upstream
conditions, if $v_{L,1} = v_{L,2} \equiv v_{{\rm shear}}$, then
$v_{{\rm drift}} = v_{{\rm shear}}$, as expected.  For the case of
symmetric reconnection with $B_1=B_2$ and $\rho_1=\rho_2$,
Eq.~(\ref{eqn::drift_eqn}) reduces to the expected result of
\begin{equation}
  v_{{\rm drift}}\sim\frac{v_{L,1} + v_{L,2}}{2},
\label{eqn::drift_eqn_sym}
\end{equation}
the average of the two upstream flow velocities (which is zero if the
flows are equal and opposite). If the densities are symmetric but the
magnetic fields are not, then
\begin{equation}
  v_{{\rm drift}}\sim\frac{B_2v_{L,1} + B_1v_{L,2}}{B_2 + B_1}
\label{eqn::drift_eqn_mag}
\end{equation}
and if the magnetic fields are symmetric but densities are not, we
find
\begin{equation}
  v_{{\rm drift}}\sim\frac{\rho_1v_{L,1} + \rho_2v_{L,2}}
  {\rho_1 + \rho_2}.
\label{eqn::drift_eqn_den}
\end{equation}
These predictions are testable with simulations.

\subsection{Prediction of the Reconnection Rate}
\label{section::rate}

We turn to the prediction of the reconnection rate of 2D asymmetric
anti-parallel reconnection with arbitrary upstream parallel flow
speeds.  The reconnection rate $E_{{\rm shear,sym}}$ for symmetric
reconnection with an equal and opposite upstream flow was recently
found to scale as \citep{Cassak11a}
\begin{equation}
  E_{{\rm shear,sym}}\sim E_0\left(1-\frac{v_{{\rm shear}}^2}{c_A^2}\right),
\label{eqn::recon_rate_sym}
\end{equation}
where $E_0$ is the reconnection rate in the absence of upstream flow
($\simeq 0.1$ for collisionless reconnection), $v_{{\rm shear}}$ is
the upstream speed of the symmetric plasma bulk flow, and $c_A$ is the
outflow speed given by the Alfv\'en speed based on the reconnecting
magnetic field.  Physically, the form of this correction is motivated
in a way similar to the suppression of reconnection by diamagnetic
effects.  \citet{Swisdak10} argued that the magnetic tension force on
the plasma due to a newly reconnected magnetic field line has to
overcome the momentum of the moving plasma in the dissipation region.

\begin{figure}[t]
\centering
\noindent\includegraphics[width=20pc]{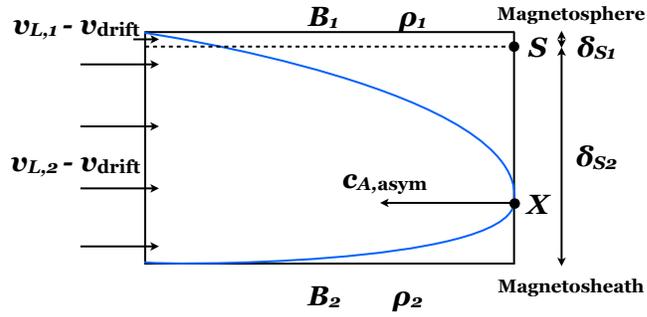}
\caption{Sketch of a newly reconnected magnetic field line (blue
  curve) for a system with upstream flow and asymmetries.  The box
  denotes the left half of a dissipation region.  ``X'' and ``S'' mark
  the X-line and stagnation point, and the dashed line goes through
  the stagnation point.  How the bulk flow impacts the dissipation
  region is sketched on the left side.  Nominally, the relative
  positions of the X-line and stagnation point are representative of
  the magnetosphere being at the top and the magnetosheath being at
  the bottom.}
\label{fig::reconnected_fieldline}
\end{figure}

For asymmetric reconnection, we propose that the effect is very
similar.  One main difference is that the outflow speed during 2D
anti-parallel asymmetric reconnection becomes \citep{Cassak07d,Swisdak07}
\begin{equation}
  c_{A,{\rm asym}}^2\sim\frac{B_1B_2}{4\pi}\frac{B_1+B_2}{\rho_1B_2 + \rho_2B_1}.
\label{eqn::alfven_outflow}
\end{equation}
A second difference is that the X-line and stagnation point are not at
the center of the dissipation region \citep{Cassak07d}, so the
upstream plasmas on the two sides contribute momentum in proportion to
the relative location of the stagnation point.  This is illustrated in
Fig.~\ref{fig::reconnected_fieldline}, showing a recently reconnected
magnetic field line in blue with the stagnation point denoted by the
``S''.  The distance from the upstream edges to the stagnation point
above and below are $\delta_{S1}$ and $\delta_{S2}$, and the left side
shows how the two upstream flows impact the straightening magnetic
field line.  Thus, not only is the upstream momentum allowed to be
different on the two sides because the density and flow speed can
differ, but they contribute in different proportions to the
dissipation region.

We follow the result of Eq.~(A3) from \citet{Swisdak10} and propose
the outflow speed decreases as
\begin{equation}
  v_{{\rm out}}^2 \sim c_{A,{\rm asym}}^2 - \frac{\delta_{S1}}{2\delta}
  (v_{L,1}-v_{{\rm drift}})^2 - \frac{\delta_{S2}}{2\delta}
  (v_{L,2}-v_{{\rm drift}})^2.
\end{equation}
In writing this, we effectively subtract off the energy due to the
flow, but here we do so in proportion to each side's contribution
opposing the release of magnetic energy.  Using $\delta_{S1}/2 \delta
= \rho_1 B_2 / (\rho_1 B_2 + \rho_2 B_1)$ and $\delta_{S2}/2 \delta =
\rho_2 B_1 / (\rho_1 B_2 + \rho_2 B_1)$ \citep{Cassak07d} and
eliminating $v_{{\rm drift}}$ using Eq.~(\ref{eqn::drift_eqn}), some
algebra reveals
\begin{equation}
  v_{{\rm out}}^2 \sim c_{A,{\rm asym}}^2 - (v_{L,1}-v_{L,2})^2 
  \frac{\rho_1 B_2 \rho_2 B_1}{(\rho_1 B_2 + \rho_2 B_1)^2}.
\end{equation}
Note, the only upstream flow dependence is on the difference in
upstream flows $v_{L,1}-v_{L,2}$.

Consequently, we suggest the reconnection rate $E_{{\rm shear,asym}}$
for asymmetric reconnection with arbitrary upstream parallel flow
speeds, extending Eq.~(\ref{eqn::recon_rate_sym}) to asymmetric
systems, scales as
\begin{equation}
  E_{{\rm shear,asym}}\sim E_{0,{\rm asym}} \left(1-\frac{v_{{\rm shear}}^2}
    {c_{A,{\rm asym}}^2} \frac{4 \rho_1 B_2 \rho_2 B_1}
    {(\rho_1 B_2 + \rho_2 B_1)^2} \right),
\label{eqn::recon_rate}
\end{equation}
where
\begin{equation}
  E_{0,{\rm asym}} \sim \frac{B_1 B_2}{B_1 + B_2} \frac{c_{A,{\rm asym}}}{c} 
  \frac{2 \delta}{L_d}
\end{equation}
is the asymmetric reconnection rate in the absence of upstream flow
\citep{Cassak07d} and
\begin{equation}
  v_{{\rm shear}} = \frac{v_{L,1} - v_{L,2}}{2}
\label{eq:vsheardef}
\end{equation}
is half the difference of the upstream flow speeds (their average if
oppositely directed).  Thus, the reconnection rate decreases with
increasing flow shear.

An important consequence of Eq.~(\ref{eqn::recon_rate}) is the
prediction of a critical flow speed $v_{{\rm shear,crit}}$ above which
reconnection does not occur (corresponding to $E_{{\rm shear,asym}} =
0$).  The prediction is that reconnection shuts off above flow shear
speeds of
\begin{equation}
  v_{{\rm shear,crit}} \sim c_{A,{\rm asym}} \frac{\rho_1 B_2 + \rho_2 B_1}
  {2 (\rho_1 B_2 \rho_2 B_1)^{1/2}}.
\label{eqn::vcrit}
\end{equation}
In the symmetric reconnection limit, this expression reduces to
$v_{{\rm shear,crit}} \sim c_A$, the known result, so it generalizes
the known result to asymmetric systems.  It is important to note that
the fraction multiplying $c_{A,{\rm asym}}$ is always greater than or
equal to one.  This implies that the critical flow shear required to
suppress asymmetric reconnection exceeds the asymmetric Alfv\'en
speed.  For larger asymmetries, the critical flow shear becomes
larger.  Consequently, while super-Alfv\'enic flow is sufficient to
suppress symmetric reconnection, reconnection can proceed in
asymmetric reconnection for super-Alfv\'enic flow, which differs from
the \citet{Cowley89} prediction.  If the asymmetry is large (as is
typical at Earth's magnetopause), then the suppression condition is
much larger than the asymmetric Alfv\'en speed.  This is discussed
further in Sec.~\ref{section::applications}.

\section{Simulation Setup}
\label{section::Setup}

To test the predictions, we perform 2D simulations using the massively
parallel two-fluid code {\sc F3D} \citep{Shay04}.  The code solves for
the density, ion velocity, magnetic field, and ion pressure (assumed
to be an adiabatic ideal gas with ratio of specific heats $\gamma =
5/3$).  Electrons are assumed cold for simplicity.  The numerical
algorithm is the trapezoidal leapfrog in time and fourth order finite
difference in space.

The code evolves variables normalized to $B_0$, the reference magnetic
field strength, and $\rho_0$, the reference mass density, where these
quantities typically are the initial values on one upstream side of
the simulation.  Other variable's normalizations are derived from
these values: velocities are normalized to the Alfv\'en speed $c_{A0}
= B_0 / (4 \pi \rho_0)^{1/2},$ lengths are normalized to the ion
inertial length $d_{i0} = (m_i c^{2} / 4 \pi n_0 e^{2})^{1/2}$, times
are normalized to the inverse ion cyclotron frequency
$\Omega_{ci}^{-1} = (e B_0 / m_i c)^{-1}$, electric fields are
normalized to $c_{A0} B_0 / c$, and pressures are normalized to $B_0^2
/ 4 \pi$, where $m_i$ and $e$ are the ion mass and charge and $n_0 =
\rho_0 / m_i$ is the plasma number density.  The $x, y$ and $z$
directions are aligned with the initial direction of the magnetic
field ($L$ in boundary normal coordinates), the inflow ($N$), and the
out-of-plane current ($M$).

Periodic boundary conditions are used in each direction.  The size of
the computational domain is $L_{x} \times L_{y} = 204.8 \times 102.4 \
d_{i0}$, and the grid scale is $0.05 \ d_{i0}$ in each direction.  The
electron mass is $m_{e} = m_{i} / 25$, and it is not expected that the
large-scale behavior seen in our simulations is sensitive to this
value.

\begin{table*}
  \caption{List of simulations with their initial 
    values of upstream magnetic fields $B_1$ and $B_2$ (in units of $B_0$), 
    mass densities $\rho_1$ and $\rho_2$ (in units of $\rho_0$), and upstream 
    flow speed $v_{{\rm shear}}$ (in units of $c_{A0}$).  
    Also included is the predicted convection speed $v_{{\rm drift,pred}}$ 
    from Eq.~(\ref{eqn::drift_eqn}), measured convection speeds of the top 
    ($v_{{\rm drift},T}$) and bottom ($v_{{\rm drift},B}$) current sheets, the 
    predicted reconnection rate 
    $E_{{\rm pred}}$ from Eq.~(\ref{eqn::recon_rate}) (in units of 
    $c_{A0} B_0 / c$), and the measured 
    reconnection rates from the top $E_T$ and bottom $E_B$ current sheets.
    The $E_{0,{\rm asym}}$ value is obtained from the average measured reconnection rate 
    in runs without shear, except for the $B_1 = 2 \ B_0$ simulation which uses the
    prediction from \citet{Cassak07d} using $\delta/L=0.06$.}
  \centering
\begin{tabular}[\textwidth]{c c c c c | c c c | c c c }
  \hline
  $B_1$ & $B_2$ & $\rho_1$ & $\rho_2$ & $v_{{\rm shear}}$ & $v_{{\rm drift,pred}}$ & $v_{{\rm drift},T}$ & $v_{{\rm drift},B}$ & $E_{{\rm pred}}$ & $E_T$ & $E_B$ \\
  \hline
  3 & 1 & 1 & 1 & 0.0 & 0.0 & 0.02 & -0.04 & 0.19    & 0.20  & 0.18 \\
  3 & 1 & 1 & 1 & 0.4 & 0.2 & 0.24 & 0.21  & 0.18    & 0.18  & 0.18 \\
  3 & 1 & 1 & 1 & 0.8 & 0.4 & 0.49 & 0.39  & 0.16    & 0.16  & 0.15 \\
  3 & 1 & 1 & 1 & 1.2 & 0.6 & 0.63 & 0.53  & 0.12   & 0.093 & 0.083 \\
  3 & 1 & 1 & 1 & 1.6 & 0.8 & 0.60 & 0.69  & 0.067   & 0.071 & 0.065 \\
  3 & 1 & 1 & 1 & 2.0 & 1.0 & 1.03 & 0.98  & 0     & ---   & --- \\
  3 & 1 & 1 & 1 & 2.4 & 1.2 & --- & --- & --- & --- & --- \\
  \hline
  2 & 1 & 1 & 1 & 1.2 & 0.4 & 0.37 & 0.37  & 0.041   & 0.049 & 0.060 \\
  \hline
  1 & 1 & 1 & 3 & 0.0 & 0.0 & ---  & ---   & 0.042   & 0.044 & 0.040 \\
  1 & 1 & 1 & 3 & 0.1 & 0.05 & ---  & ---   & 0.041   & 0.037 & 0.038 \\
  1 & 1 & 1 & 3 & 0.2 & 0.1 & ---  & ---   & 0.039   & 0.039 & 0.039 \\
  1 & 1 & 1 & 3 & 0.4 & 0.2 & ---  & ---   & 0.032   & 0.036 & 0.036 \\
  1 & 1 & 1 & 3 & 0.6 & 0.3 & ---  & ---   & 0.019   & 0.027 & 0.025 \\
  1 & 1 & 1 & 3 & 0.8 & 0.4 & ---  & ---   & 0.0017    & ---   & --- \\
  \hline
\end{tabular}
\label{table::runs}
\end{table*}

The initial conditions for the simulation have a magnetic field
profile given by an asymmetric double Harris sheet,
\begin{equation}
  B_{x}(y) = \left\{ \begin{array}{ll} -B_{01}
      \tanh\left(\frac{|y|-L_{y}/4}{w_{0}}\right) & \hskip 0.1cm 
      L_{y} / 4 < |y| < L_{y} /2 \\ 
      -B_{02} \tanh\left(\frac{|y|-L_{y}/4}{w_{0}}\right) & \hskip 0.52cm 
      0 < |y| < L_{y}/4  \end{array} \right. 
\end{equation}
with an initial current sheet width of $w_{0} = d_{i0}$.  We do not
use an initial out-of-plane (guide) magnetic field.  The initial
density profile is
\begin{equation}
  \rho(y) = \frac{\rho_{01} + \rho_{02}}{2} - \frac{\rho_{01} -
    \rho_{02}}{2} \tanh\left(\frac{|y|-L_{y}/4}{w_{0}}\right),
\label{initrho}
\end{equation}
with asymptotic values $\rho_{01}$ in the central portion of the
domain and $\rho_{02}$ at the top and bottom of the domain.  The
initial pressure profile is chosen to balance the pressure
identically, with a minimum value of $\beta_{{\rm min}} B_{{\rm
    max}}^2/8 \pi$, where $B_{{\rm max}} = \max(B_{01},B_{02})$ is the
stronger of the two magnetic field strengths and $\beta_{{\rm min}}$
is the minimum plasma beta on either side of the plasma.  We choose
$\beta_{{\rm min}} = 2$.  

The bulk flow is initialized with a profile of
\begin{equation}
  v_{x}(y) = \left\{ \begin{array}{ll} -v_{1}
      \tanh\left(\frac{|y|-L_{y}/4}{w_{0}}\right) & \hskip 0.1cm 
      L_{y} / 4 < |y| < L_{y} /2 \\ 
      -v_{2} \tanh\left(\frac{|y|-L_{y}/4}{w_{0}}\right) & \hskip 0.52cm 
      0 < |y| < L_{y}/4.  \end{array} \right. 
\end{equation}
We use the same $w_{0}$ as for the magnetic field profile (although
this is not a requirement \citep{Li10}).  The upstream flow speeds are
$v_1$ and $v_2$, which for the simulations presented here, are always
equal and opposite: $v_1 = -v_2$.  We define this common speed as
$v_{{\rm shear}}$, which points to the left at the top and bottom of
the computational domain and to the right in the central part of the
domain.

The simulations do not employ any explicit resistivity or viscosity.
However, there is a fourth order diffusion used in each evolution
equation to damp noise at the grid scale, with coefficient $1 \times
10^{-4}$ for asymmetric magnetic field simulations and $1.25 \times
10^{-5}$ for asymmetric density simulations.  The magnetic field is
perturbed initially with random noise of amplitude $0.00005 \ B_0$ to
break symmetry, which allows any secondary magnetic islands that arise
to be ejected.  A coherent magnetic perturbation of the form
$\delta{\bf B} = -(0.012 B_0 L_{y} / 2 \pi) \hat{{\bf z}} \times
\nabla [\sin(2 \pi x/ L_{x}) \sin^{2}(2 \pi y/L_{y})]$ is used to
initiate reconnection in a controllable manner.

A series of numerical simulations is performed with initial magnetic
asymmetries of 3 or 2 with uniform density and with plasma density
asymmetries of 3 with symmetric magnetic field.  The initial flow
speed $v_{{\rm shear}}$ in the upstream regions is varied between 0
and $2.4 \ c_{A0}$.  Table~\ref{table::runs} shows a list of the
various simulations from which data are collected.

\section{Results}
\label{section::Results}

\begin{figure}
\centering
\noindent\includegraphics[width=20pc]{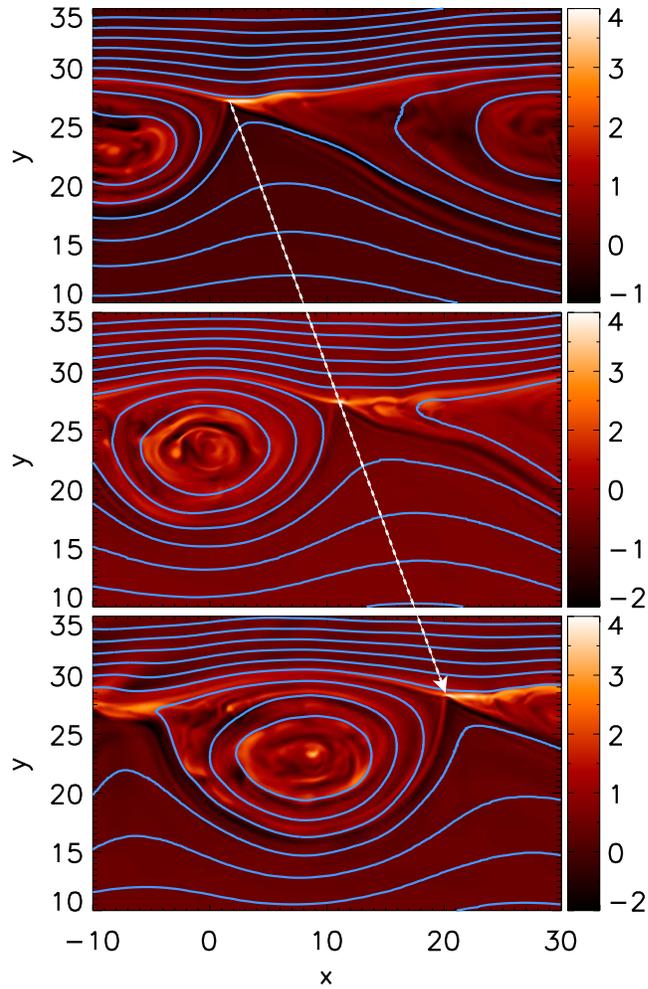}
\caption{Out-of-plane current density $J_z$ as a function of $x$ and
  $y$ (in units of $d_{i0}$), with magnetic field lines overplotted in
  blue, for a simulation with $v_{{\rm shear}} = 1.2 \ c_{A0}$, $B_1=3
  \ B_0, B_2= B_0$, and $\rho_1 = \rho_2 = \rho_0$.  The top, middle,
  and bottom plots are at time $t = 90, 105,$ and $120 \
  \Omega_{ci}^{-1}$.  Only a portion of the computational domain is
  plotted.}
\label{fig::overview}
\end{figure}

We begin with a case study showing that the X-line in asymmetric
reconnection convects in the outflow direction for equal and opposite
upstream flow speeds.  The simulation has asymmetric magnetic fields
$B_1 = 3 \ B_0$ and $B_2 = B_0$ and symmetric density $\rho_1 = \rho_2 =
\rho_0$ with an upstream flow speed of $v_{{\rm shear}} = 1.2 \
c_{A0}$.  Figure~\ref{fig::overview} shows the out-of-plane current
density $J_z$ as the background, with magnetic field lines overplotted
in blue for a portion of the computational domain.  The top, middle,
and lower panels are at times $t = 90, 105$, and $120 \
\Omega_{ci}^{-1}$, respectively.  The X-line, located near the peak in
the current density, clearly convects to the right as shown by the
white arrow.  It travels approximately 18 $d_{i0}$ in 30
$\Omega_{ci}^{-1}$, giving a convection speed close to $0.6 \ c_{A0}$,
which is the predicted speed from Eq.~(\ref{eqn::drift_eqn}).

To extract more quantitative information from the simulations, the
systems are evolved until a steady-state is reached with the X-line
convection speeds and reconnection rates being relatively constant.
To apply a common criterion for the steady-state across simulations
(which have different reconnection rates), we put a condition on the
size of the primary magnetic islands; all simulations presented here
are relatively steady when the island half widths are between 7 and
$11 \ d_{i0}$.  X-line convection speeds and reconnection rates are
measured during these times.

The X-line and O-line are found in the standard way as the saddle
point and local extremum, respectively, near the current sheets of the
flux function $\psi$ given by ${\bf B} = {\bf \hat{z}} \times \nabla
\psi$.  The X-line convection speed is obtained from the time average
of the time derivative of the position of the X-line during the
steady-state interval.  The reconnection rate is the average of the
time rate of change of the difference in magnetic flux between the
X-line and the O-line.

\subsection{The X-line Convection Speed}
\label{section::Results_convection}

\begin{figure}
\centering
\noindent\includegraphics[width=20pc]{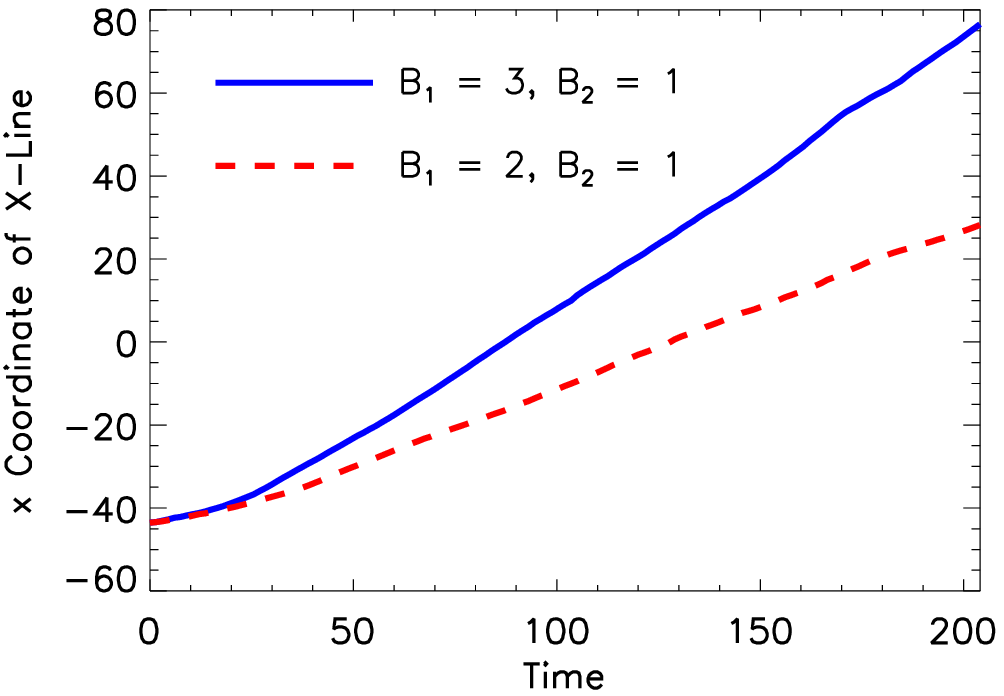}
\caption{X-line position (in units of $d_{i0}$) as a function of time
  (in units of $\Omega_{ci}^{-1}$) for simulations with $v_{{\rm
      shear}} = 1.2 \ c_{A0}$ for $B_1=3 \ B_0, B_2=B_0$ (blue solid)
  and $B_1=2 \ B_0, B_2= B_0$ (red dashed).  The average drift speeds
  are $v_{{\rm drift}}=0.63 \ c_{A0}$ for $B_1 = 3 \ B_0$ and $v_{{\rm
      drift}}=0.37 \ c_{A0}$ for $B_1 = 2 \ B_0$.}
\label{fig::Drift_Asym}
\end{figure}

For the X-line convection speed, we first use simulations with
asymmetric magnetic fields, symmetric plasma densities, and symmetric
upstream flow.  The $x$ coordinate of the X-line in the simulation
reference frame is plotted as a function of time in
Fig.~\ref{fig::Drift_Asym}.  The (blue) solid line is from a
simulation with $B_1=3 \ B_0, B_2= B_0$ and the (red) dashed line is
for $B_1=2 \ B_0, B_2=B_0$, both with $v_{{\rm shear}}=1.2 \ c_{A0}$.
The convection speed is higher with the stronger magnetic field, as
predicted, with measured values of $v_{{\rm drift}} = 0.63 \ c_{A0}$
for $B_1 = 3 \ B_0$ and $0.37 \ c_{A0}$ for $B_1 = 2 \ B_0$, compared
to predicted values from Eq.~(\ref{eqn::drift_eqn_mag}) of 0.60 and
$0.40 \ c_{A0}$.

\begin{figure}
\centering
\noindent\includegraphics[width=20pc]{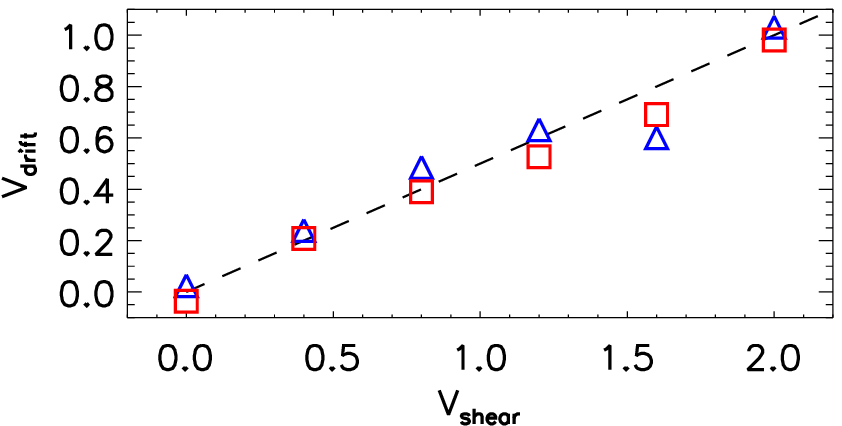}
\caption{X-line convection speed $v_{{\rm drift}}$ (in units of
  $c_{A0}$) for the top ($\triangle$) and bottom ($\square$) current
  sheets from simulations with magnetic fields $B_1=3 \ B_0$ and
  $B_2=B_0$ as a function of upstream flow speed $v_{{\rm shear}}$ (in
  units of $c_{A0}$).  The dashed line is the prediction from
  Eq.~(\ref{eqn::drift_eqn_mag}).}
\label{fig::Drift_Flow}
\end{figure}

To test the upstream flow speed dependence, Fig.~\ref{fig::Drift_Flow}
shows the average drift speed $v_{{\rm drift}}$ as a function of
upstream flow speed $v_{{\rm shear}}$ for simulations with $B_1 = 3 \
B_0$ and $B_2 = B_0$.  Here, and throughout, the (blue) triangles are
for the top current sheet and the (red) squares are for the bottom
current sheet.  The prediction from Eq.~(\ref{eqn::drift_eqn_mag}) is
plotted as the dashed line, and the results clearly agree well with
the prediction.

Simulations with symmetric magnetic fields but with asymmetric plasma
densities were also carried out.  However, we do not expect the
numerical results to be reliable because of a known problem with the
fluid approach in systems with asymmetric density.  In particular,
fluid simulations require conduction to allow for plasma mixing
\citep{Cassak09c}; newly reconnected field lines have different
densities and temperatures, and in the absence of conduction the
strong parallel temperature gradient persists instead of the plasmas
mixing.  These simulations do not contain conduction and it is
unlikely the standard fluid closure reproduces the more realistic
kinetic mixing in a collisionless dissipation region.
Particle-in-cell simulations will be required to assess this
prediction for asymmetric densities.

\subsection{The Reconnection Rate}
\label{section::Results_rrate}

\begin{figure}
\centering
\noindent\includegraphics[width=20pc]{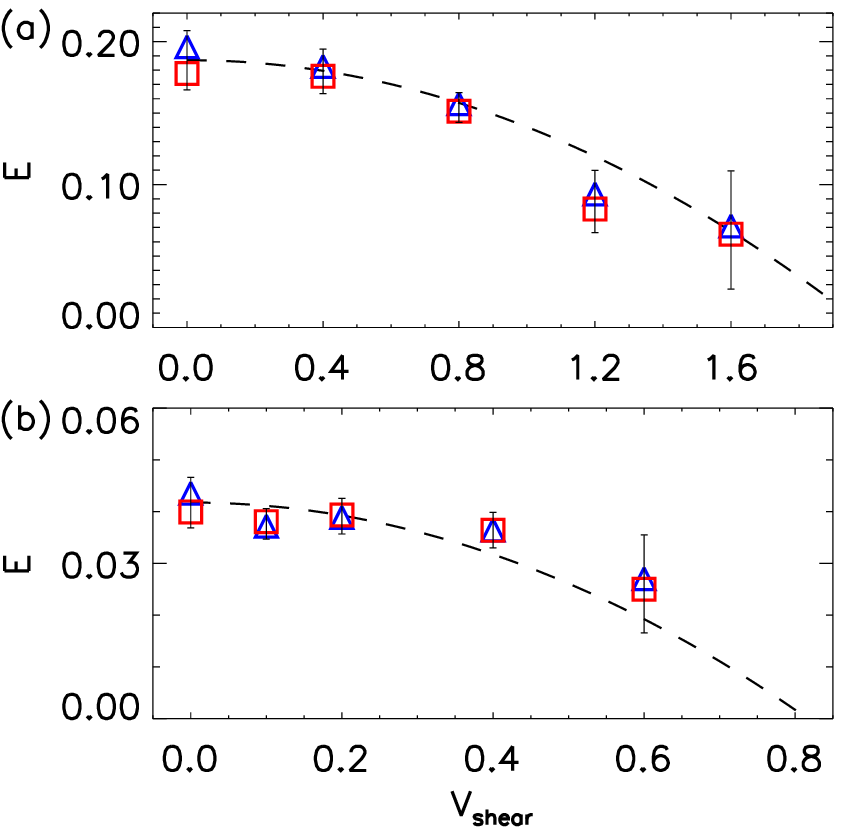}
\caption{Reconnection rate $E$ (in units of $c_{A0} B_0 / c$) for the
  top ($\triangle$) and bottom ($\square$) current sheets from
  simulations with various flow speed $v_{{\rm shear}}$ (in units of
  $c_{A0}$) and (a) magnetic field asymmetry $B_1=3 \ B_0$ and $B_2=
  B_0$ and (b) density asymmetry $\rho_1=\rho_0$ and $\rho_2 = 3 \
  \rho_0$.  The dashed lines are the predicted rates from
  Eq.~(\ref{eqn::recon_rate}), with $E_{0,{\rm asym}}$ as the average
  rate of reconnection in the absence of a sheared flow.}
\label{fig::Rate_Flow}
\end{figure}

For the reconnection rate in systems with asymmetries and upstream
flow, we plot the reconnection rate $E$ as a function of flow speed
$v_{{\rm shear}}$ in Fig.~\ref{fig::Rate_Flow}(a) for simulations with
fixed $B_1 = 3 \ B_0$ and $B_2 = B_0$ for upstream flow speeds from 0
to $1.6 \ c_{A0}$.  The prediction from Eq.~(\ref{eqn::recon_rate}) is
plotted as the dashed line, where $E_{0,{\rm asym}}$ is chosen as the
measured value of the reconnection rate in the simulation with no
upstream flow.  The agreement is very good.  The result for $v_{{\rm
    shear}} = 1.2 \ c_{A0}$ falls a little below the curve; this
simulation had a significant secondary island unlike the others which
motivates its slightly worse performance.

A similar study is carried out with simulations of asymmetric plasma
densities but with symmetric magnetic field.  While the convection
speed is not expected to be correct in these simulations, it was
argued that the reconnection rate is reliable
\citep{Cassak09c,Birn10}.  (This is because redoing the calculation of
the outflow speed and reconnection rate from \citet{Cassak07d} for a
system where the downstream plasma does not mix leads to the same
expressions as for the system where mixing occurs.  The differences
arise only in the sub-structure of the dissipation region.)  The
results for a simulation study with $\rho_1 = \rho_0$ and $\rho_2 = 3
\ \rho_0$ with upstream flow speeds varied from 0 to $0.6 \ c_{A0}$
are shown in Fig.~\ref{fig::Rate_Flow}(b).  The prediction from
Eq.~(\ref{eqn::recon_rate}) is plotted as the dashed line.  For
asymmetric densities, the agreement with the prediction is again very
good.

\subsection{The Cessation Condition}
\label{section::Results_cessation}

\begin{figure}
\centering
\noindent\includegraphics[width=20pc]{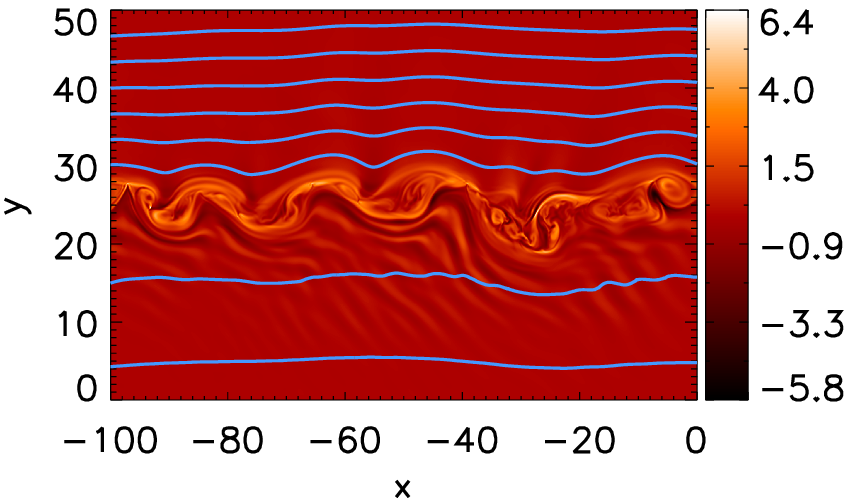}
\caption{Out-of-plane current density $J_z$, with magnetic field lines
  overplotted in blue, for a simulation with $B_1 = 3 \ B_0, B_2 =
  B_0, \rho_1 = \rho_2 = \rho_0$ with $v_{{\rm shear}} = 2.4 \
  c_{A0}$, which is above the predicted cutoff where reconnection is
  suppressed.}
\label{fig::cessation}
\end{figure}

We now test the condition for the cessation of reconnection, given by
Eq.~(\ref{eqn::vcrit}).  For the series of simulations with $B_1 = 3 \
B_0, B_2 = B_0, \rho_1 = \rho_2 = \rho_0$, we have $c_{A,{\rm asym}} =
\sqrt{3} \ c_{A0}$ from Eq.~(\ref{eqn::alfven_outflow}), so $v_{{\rm
    shear,crit}} = 2 \ c_{A0}$.  In the simulations, we see that
reconnection occurs, with the predicted reconnection rate and X-line
drift speed, for $v_{{\rm shear}} = 1.6 \ c_{A0}$.  In contrast, for
$v_{{\rm shear}} = 2.4 \ c_{A0}$, there is a phase change and
reconnection is not the dominant effect.  This is shown in
Fig.~\ref{fig::cessation}, which gives the out-of-plane current
density $J_z$ for this simulation.  There is clear evidence of the
early phases of a Kelvin-Helmholtz instability.  We note that
reconnection is known to occur as a secondary instability of
Kelvin-Helmholtz in a plasma, and that does occur here, but this is
fundamentally different than the reconnection-dominated current sheets
displayed in Fig.~\ref{fig::overview}.  We note that for $v_{{\rm
    shear}} = 2 \ c_{A0}$, right at the predicted cutoff, there is a
hybrid of reconnection and Kelvin-Helmholtz rolls forming.  Thus, the
results are consistent with the prediction.  As the qualitative
behavior near the cutoff becomes more challenging to assess, it is
prohibitive to pin down the transition flow speed to higher precision
than done here.

The asymmetric density simulations with $B_1 = B_2 = B_0, \rho_1 =
\rho_0,$ and $\rho_2 = 3 \ \rho_0$ give another opportunity to test
the prediction.  Here, $c_{A,{\rm asym}} = \sqrt{1/2} \ c_{A0}$, so
$v_{{\rm shear,crit}} = \sqrt{2/3} \ c_{A0} \simeq 0.82 \ c_{A0}$.  In
the simulations, reconnection does occur for $v_{{\rm shear}} = 0.6 \
c_{A0}$ but is not appreciable for $v_{{\rm shear}} = 0.8 \ c_{A0}$,
which is consistent with the predictions.

In summary, the predictions for $v_{{\rm drift}}$ in
Eq.~(\ref{eqn::drift_eqn}), $E_{{\rm shear,asym}}$ in
Eq.~(\ref{eqn::recon_rate}), and $v_{{\rm shear,crit}}$ in
Eq.~(\ref{eqn::vcrit}) agree with the simulation results.

\section{Applications to Planetary Magnetospheres}
\label{section::applications}

For a sample application, we consider reconnection at the cusps of
Earth's magnetosphere.  We emphasize that the predictions in
Sec.~\ref{section::Theory} assume an isolated X-line; we first discuss
how the present results suggest isolated X-lines would act in the
magnetosphere.  We call side 1 ``ms'' for magnetosphere and side 2
``sh'' for magnetosheath.  For typical conditions at the magnetopause,
$\rho_{sh} \gg \rho_{ms}$ and $B_{ms} \gtrsim B_{sh}$.  Also, the
upstream flow speed in the magnetosphere $v_{L,ms}$ is negligible,
while the magnetosheath flow $v_{L,sh}$ is due to the solar wind
flowing around the magnetopause.  In this limit, the X-line convection
speed from Eq.~(\ref{eqn::drift_eqn}) becomes
\begin{equation}
  v_{{\rm drift}}\simeq v_{L,sh}. \label{equation::driftms}
\end{equation}
Thus, an isolated X-line would convect tailward at essentially the
same speed as the flow in the magnetosheath, not the average flow
speed as one would expect for symmetric reconnection.  The physical
cause for this, as sketched in Fig.~\ref{fig::reconnected_fieldline},
is that the stagnation point is far to the magnetospheric side of the
dissipation region, so most of the dissipation region is populated by
magnetosheath plasma and most of its momentum is contained in the
magnetosheath plasma.  Therefore, it flows near the speed of the
magnetosheath plasma.

This is borne out in recent Cluster observations of a tailward
convecting X-line \citep{Wilder14} which we use as a case study.  In
this event at the southern hemisphere's cusp on December 27th, 2005,
Cluster's C1 and C3 spacecraft witnessed reconnection signatures on a
crossing from the magnetosheath to the magnetosphere.  The
magnetosheath had $B_{sh} \simeq 10-15$ nT and a number density
$n_{sh} \simeq 60 - 70$ cm$^{-3}$, while in the magnetosphere $B_{ms}
\simeq 60$ nT and $n_{ms} = 0.5$ cm$^{-3}$.  From their separation and
the time delay between observed jet reversals, the convection speed of
the X-line was estimated to be $v_{{\rm drift}} =$ 105 km/s, while the
$L$ component of the magnetosheath flow was estimated at $v_{L,sh} =
106$ km/s.  This is consistent with Eq.~(\ref{equation::driftms}).

In the past, assessing whether reconnection could occur was done by
comparing the magnetosheath flow speed to the magnetosheath Alfv\'en
speed \citep{Cowley89}.  For this event, the magnetosheath Alfv\'en
speed is 28 km/s, while the magnetosheath flow speed is 105 km/s.
This exceeds twice the magnetosheath Alfv\'en speed so previous models
would suggest reconnection should not occur, but reconnection is
observed to happen.  The asymmetric Alfv\'en speed for these
parameters is 74.5 km/s, so $v_{{\rm shear}} = v_{L,sh} / 2 \simeq$ 53
km/s is sub-Alfv\'enic, and the prediction here is that reconnection
should occur.

For the reconnection rate, in the limit appropriate for typical
conditions at the magnetopause, Eq.~(\ref{eqn::recon_rate}) becomes
\begin{equation}
  E_{{\rm shear,asym}}\sim E_{0,{\rm asym}} \left(1-\frac{4 v_{{\rm shear}}^2}
    {c_{A,{\rm asym}}^2} \frac{\rho_{ms} B_{sh}}
    {\rho_{sh} B_{ms}} \right).
\label{eqn::reconratemsp}
\end{equation}
Since $\rho_{sh} \gg \rho_{ms}$, this implies the surprising result
that the reconnection rate is not changed much by the flow shear for
typical magnetospheric parameters.  This is a major departure from the
symmetric case where the reconnection rate falls strongly as a
function of flow shear speed.  Physically, the cause for this is
related to the stagnation point being very close to the magnetospheric
side, as before.  In the reference frame of the X-line, the
magnetosheath is essentially still, while the magnetosphere has a
significant flow.  However, since the magnetospheric density is so
low, it has very little effect on the reconnection process.

This is seen further by investigating the condition for suppressing
reconnection.  From Eq.~(\ref{eqn::reconratemsp}) and using $v_{{\rm
    shear}} \simeq v_{L,sh}/2$ from Eq.~(\ref{eq:vsheardef}) since the
magnetospheric plasma is essentially stationary, the condition that
reconnection is suppressed ($E_{{\rm shear,asym}} \leq 0$) is
\begin{equation}
  v_{L,sh} \geq c_{A,{\rm asym}} \left(\frac{\rho_{sh} B_{ms}}
  {\rho_{ms} B_{sh}}\right)^{1/2}.
\label{eqn::vcritmsp}
\end{equation}
For the example of the case study event from Cluster, this implies it
would take a magnetosheath flow speed of 22 times bigger than
$c_{A,{\rm asym}} \simeq [B_{sh} (B_{ms}+B_{sh}) /
  4\pi\rho_{sh}]^{1/2}$ for flow shear to suppress reconnection, far
greater than in symmetric reconnection and the \citet{Cowley89} model
where Alfv\'enic flow suppresses reconnection.  (We point out further
that particle-in-cell simulations and observations typically find
outflow speeds half of that predicted in symmetric or asymmetric
reconnection theories, so the critical speed may be a factor of two
smaller than this prediction.  However, this does not change the
result that the critical speed is much higher than the asymmetric
Alfv\'en speed and the magnetosheath Alfv\'en speed.)  Thus, for an
isolated X-line with typical magnetospheric parameters, the flow shear
would rarely prevent reconnection from occurring.

While X-lines in the magnetosphere can convect as discussed earlier,
it was suggested from observations \citep{Fuselier00b,Frey03} that
cusp reconnection can be steady for hours with a stationary X-line,
though it may be difficult to be sure solely from auroral signatures.
Further, the distance to the X-line has been inferred from
observations \citep{Petrinec03,Trattner07a,Trattner07b}, and strong
convection is not seen, even though Eq.~(\ref{equation::driftms})
would predict a large enough convection speed to be observable.  This
suggests that reconnection in the magnetosphere is not always
consistent with isolated X-lines, which was a key assumption of the
theory presented here.  In particular, the theory does not take any
external effect other than flow into account.  A single dayside X-line
is not isolated; instead, for a single X-line during steady times to
be stationary, it must balance all magnetic forces, gas pressure
forces, and dynamic pressure forces from bulk flow (of the type
discussed here).  Thus, by definition, the flow effects have already
been taken into account to set up a single X-line, and the present
model would not apply.  Physically, the present model does not include
magnetic fields in the cusp being line-tied to the ionosphere, which
is an additional force that can slow or prevent X-line convection and
would undoubtedly modify the present results.  It is possible that the
tailward moving X-lines seen in observations
\citep{Hasegawa08,Wilder14} are related to a secondary island moving
tailward; a secondary X-line is not in the same equilibrium as a
single X-line, so it can convect tailward.  Future work is necessary
to determine the conditions under which some X-lines convect and
others do not, the speed at which they convect, and their signatures
in the cusp.

A second point worth making is that the results in this section would
change in the presence of a high-density plasmaspheric drainage plume
\citep{Borovsky06a,Borovsky06b} on the magnetospheric side of the
dissipation region.  This increases the local density, which decreases
the outflow speed \citep{Walsh14,Walsh14b}.  Plumes are most commonly
observed centered around 13.6~h magnetic local time \citep{Walsh13}.
If the magnetospheric density is comparable or larger than the
magnetosheath, the X-line would not convect with the magnetosheath
flow, and the full expressions from Sec.~\ref{section::Theory} would
be needed.

We note in passing that the significant departures between symmetric
and asymmetric reconnection with a flow shear suggest that the effect
of flow shear at Jupiter, Saturn, and Uranus should be revisited.  The
previous work on this subject employed the predictions for symmetric
reconnection, so the results are likely to change with the asymmetric
reconnection results presented here.

It is worth putting the present results in context of a leading model
for what allows cusp reconnection to occur despite nominally
super-Alfv\'enic flow.  It was argued that the draped magnetic field
during northward IMF compresses, and therefore increases, the
magnetopause magnetic field, which itself causes the mass density to
decrease, forming a density depletion layer at the magnetopause
\citep{Petrinec03}.  Both the increased field and decreased density
increase the local Alfv\'en speed, potentially making the flow
sub-Alfv\'enic.  The present theory {\it should not} be interpreted as
implying that a density depletion layer does not occur or that the
layer would not change the local Alfv\'en speed.  Using the
\citet{Wilder14} parameters, \citet{Petrinec03} would suggest
(assuming that the Alfv\'en Mach number is based on magnetosheath
parameters) that a density depletion by a factor of about 2.5 would be
sufficient to make the flow sub-Alfv\'enic (see their Fig.~7).  When
based on the asymmetric Alfv\'en speed, the necessary depletion factor
reduces to approximately 1.3.  Clearly, more research is needed to
identify the level of plasma depletion, its effect on cusp
reconnection, and the extent to which the present results apply to the
dayside magnetopause.

\section{Discussion}
\label{section::Discussion}

In this study, we use a scaling argument based on conservation of
momentum to find the convection (drift) speed of the X-line in
anti-parallel 2D asymmetric magnetic reconnection with arbitrary
upstream parallel flow speeds [Eq.~(\ref{eqn::drift_eqn})].  We also
present a prediction for the reconnection rate for arbitrary upstream
flow speeds [Eq.~(\ref{eqn::recon_rate})].  The predictions are
confirmed using 2D two-fluid numerical simulations.  For asymmetric
magnetic fields, the results agree well for both predictions.  For
asymmetric densities, simulations agree with the reconnection rate
prediction, but the drift speed cannot be assessed because the fluid
model does not correctly model plasma mixing.  In particular, the
results show that the X-line convects even for sub-Alfv\'enic flow,
which contrasts the leading model by \citet{Cowley89}.

The reconnection rate prediction gives a threshold flow shear above
which reconnection does not occur [Eq.~(\ref{eqn::vcrit})], and this
prediction is consistent with the simulations.  This result shows that
asymmetric reconnection can persist with flow shear exceeding the
asymmetric Alfv\'en speed, and can occur even with much larger flow
shear speeds if the asymmetries are large.  We note that the critical
speed prediction in Eq.~(\ref{eqn::vcrit}) differs from the condition
given by \citet{LaBelleHamer95}, who suggested the critical speed is
the larger of the two Alfv\'en speeds on either upstream side of the
dissipation region.  It also differs from the prediction that
reconnection does not occur if the difference between the
magnetosheath and magnetospheric flow exceeds twice the magnetosheath
Alfv\'en speed \citep{Cowley89}.  To see this, consider the simulation
with $B_1 = 3, B_2 = 1$, a uniform density of $\rho_1 = \rho_2 = 1$,
and a flow of $v_{{\rm shear}} = 1.6$.  Here, the magnetosheath
nominally corresponds to the ``2'' side (with the weaker magnetic
field).  In the rest frame of the magnetosphere (the ``1'' side), the
magnetosheath flow speed is 3.2.  This exceeds double the
magnetosheath Alfv\'en speed of 1.  The fact that reconnection is
observed in this simulation is evidence against the \citet{Cowley89}
model.  The fact that we do not see reconnection in the $v_{{\rm
    shear}} = 2.4$ simulation is evidence against the
\citet{LaBelleHamer95} model, which is below the magnetospheric
Alfv\'en speed of 3.

The results have potentially important implications for reconnection
at Earth's magnetopause.  For isolated X-lines, the predictions
suggest that the X-line convects essentially with the magnetosheath
flow as a consequence of the stagnation point being nearly all the way
to the magnetospheric side of the dissipation region.  The
reconnection rate is affected by only a small amount, and it would
take magnetosheath flow an order of magnitude faster than the
asymmetric Alfv\'en speed to suppress reconnection.  This is a major
departure from the current understanding based on symmetric
reconnection which claims magnetosheath flow greater than twice the
Alfv\'en speed is sufficient to suppress reconnection.  The results
may differ for non-isolated X-lines, such as magnetic fields line-tied
to the ionosphere during reconnection near the polar cusps.  The
results could also drastically alter previous estimates of how flow
shear affects dayside reconnection at Jupiter, Saturn, and Uranus, so
revisiting these results is important future work.  Another potential
application is determining whether the present results impact
predictions for the speed of anti-sunward propagation of flux transfer
events \citep{Cowley89,Cooling01}.

The present analysis made a number of simplifying assumptions that
should be addressed in future work.  Since the analytical calculation
uses the fluid picture, finite Larmor radius effects are ignored, but
they may be important in the boundary layers \citep{Malakit13,Koga14},
especially for the dayside magnetopause where the density asymmetry is
typically significant.  The present analysis and simulations ignore
asymmetries in the outflow direction \citep{Murphy10,Oka11}, which may
be important for reconnection near the polar cusps.  In particular,
asymmetries in the outflow direction cause the outflow speeds to be
different in the two outflow directions \citep{Murphy10}, so future
work is necessary to determine the effect of shear flow in such
systems.  The present analysis and simulations also do not include a
guide field, which may be present \citep{Muzamil14}.  The guide field
is important because, when coupled with a gas pressure gradient across
the inflow direction at the dissipation region, it can set up
diamagnetic effects which also have been shown to cause the X-line to
convect in the outflow direction \citep{Swisdak03,Beidler11}, and
therefore either reinforce or oppose the convection caused by flow
shear \citep{Tanaka10}.  The analysis and simulations also do not
include upstream flows in the out-of-plane direction
\citep{Wang12,Chen13,Wang14,Tassi14}.  Also, the simulations used the
fluid model, which does not self-consistently capture plasma mixing in
the exhaust for systems with asymmetric density \citep{Cassak09c};
this would be fixed using simulations employing the particle-in-cell
technique \citep{Roytershteyn08,Tanaka10}.  Another future study
should address asymmetric reconnection with a flow shear in line-tied
systems relevant to polar cusp applications.

\begin{acknowledgments}
  Support from NASA West Virginia Space Grant Consortium (CED and
  CMK), the West Virginia University Summer Undergraduate Research
  Experience (SURE) program (CED), and NSF Grant AGS-0953463 (PAC) is
  gratefully acknowledged.  PAC acknowledges support from the
  International Space Science Institute in Bern, Switzerland.  This
  research used resources of the National Energy Research Scientific
  Computing Center, a DOE Office of Science User Facility supported by
  the Office of Science of the U.S. Department of Energy under
  Contract No. DE-AC02-05CH11231.  The data used to produce the
  results of this paper are available from the authors.  We thank
  M.~T.~Beidler, R.~C.~Fear, S.~A.~Fuselier, and B.~M.~Walsh for
  helpful conversations.
\end{acknowledgments}

\vspace*{0.5cm}

\noindent (a) Current address: NASA/Goddard Space Flight Center,
Greenbelt, MD, USA


\end{article}
\end{document}